\documentclass[twocolumn,showpacs,preprintnumbers,amsmath,amssymb,prl]{revtex4}
\usepackage{graphicx}
\usepackage{textcomp}
\usepackage{nicefrac}

\newcommand{\vc}[1]{\mathbf #1}

\begin{document}

\author{Heinrich Stolz}
\author{Dirk Semkat}
\affiliation{Institut f\"ur Physik, Universit\"at Rostock, D-18501 Rostock, Germany}
\title{Unique signatures for Bose-Einstein condensation in the decay luminescence lineshape of weakly interacting excitons in a potential trap}

\date{\today}

\begin{abstract}
We calculate the spatially resolved optical emission spectrum of a weakly interacting Bose gas of excitons confined in a three dimensional potential trap due to interband transitions involving weak direct and phonon mediated exciton-photon interactions. Applying the local density approximation, we show that for a non-condensed system the spatio-spectral lineshape of the direct process reflects directly the shape of the potential. The existence of a Bose-Einstein condensate changes the spectrum in a characteristic way so that it directly reflects the constant chemical potential of the excitons and the renormalization of the quasiparticle excitation spectrum. Typical examples are given for parameters of the lowest yellow excitons in $\mathrm{Cu_2O}$.
\end{abstract}

\pacs{78.20.-e,78.30.-j,71.35.Lk}

\keywords{excitons, Bose-Einstein condensation, potential traps}

\maketitle
The condensation of bosons into the system ground state at sufficiently low temperature in thermal equilibrium is one of the manifestations of quantum nature of matter \cite{einstein1925}. To reach a sufficient density, the concept of trapping the particles in a potential well has allowed the realization of atomic Bose-Einstein condensates \cite{wiemann1995,ketterle1995}. Also for bosonic quasiparticles, like microcavity polaritons, this concept has been fruitful \cite{snoke2007}. For excitons, bound electron-hole-pair excitations in semiconductors, which have been the first type of quasiparticles where Bose-Einstein condensation (BEC) has been predicted (for an overview see \cite{snoke2000}), the use of potential traps has a long history. Especially promising have been the exciton states in the semiconductor cuprous oxide (Cu$_2$O). Due to their optically forbidden nature, long lifetime are expected and one should be able to trap a very large number of particles in quasi thermal equilibrium (see, e.g., \cite{wolfe1986,snoke2000b}). 
However, despite several experimental studies of dense exciton states in this material \cite{wolfe1986,wolfe1993,snoke1990,snoke2000,naka2002}, none of these resulted in a clear demonstration of the existence of a Bose condensed state of excitons. We will show in this paper that a reason for these failures might be a wrong assignment of the decay luminescence spectrum of an exciton condensate, since all earlier papers used only qualitative arguments where the condensate has been put by hand into the spectrum, but a rigorous calculation of the decay luminescence spectrum of excitons in a trap under weak exciton-photon interaction has not been performed. 
In this paper we will give a calculation based on a mean field description of the exciton system, which not only clarifies these issues but predicts significant changes in the luminescence spectrum in the presence of a condensate. This will provide unique criteria for the onset of BEC in an excitonic system.

In the theory of interacting Bose gases, several approximations have been developed \cite{griffin1996,bergeman2000}, for a review see \cite{pethick}. The critical temperature for BEC can be roughly approximated by that of a noninteracting system.
Taking the confining potential to be that of a 3d harmonic oscillator $V_{\rm ext}= \alpha r^2$, it is given by
\begin{equation}
k_BT_{c0} =\hbar \Omega_0 \left( \frac{N}{\zeta(3)}\right)^{\nicefrac{1}{3}}
\end{equation} 
with $\Omega_0= \sqrt{2\alpha/M}$ being the oscillator frequency, $N$ the total number of particles in the trap, $M$ the mass of the particles, and $\zeta$ the Riemann zeta function. For an exciton mass of $M=2.6 m_e$ that represents the mass of the paraexcitons in Cu$_2$O \cite{brandt2007} and typical $\alpha$ parameters of exciton traps ($\alpha \simeq 0.05$ \textmu eV/\textmu m$^2$, see, e.g., \cite{wolfe1986}), this results in  $\hbar\Omega_0= 72$ neV and for $N=10^{10}$ excitons in the trap a critical temperature of 1.27 K is expected. For the noninteracting system, in the condensate all particles are in the ground state of the oscillator the size of which is $a_{osc}=\sqrt{\hbar/M\Omega_0}= 0.74$ \textmu m. From this it was concluded in previous investigations that the spatial shrinking of the luminescence line is an indication of the onset of BEC \cite{wolfe1986,snoke2000}. However, for an interacting system this is not the case. Here the condensate will form a cloud with radius $R_0= \sqrt{15}\left( N a_S/a_{osc} \right)^{\nicefrac{1}{5}}$ where $a_S$ is the s-wave scattering length \cite{pethick}, which for large $N$ can be much larger than $a_{osc}$.  Assuming again $N=10^{10}$ particles, a temperature of $T= 0.5$ K, and a scattering length of $2.8 a_B$ \cite{ohara2000}, the size of the cloud is $R_0=39$ \textmu m, much larger than the size of the oscillator ground state and also  the de Broglie wavelength $\lambda_B= \sqrt{2\pi \hbar^2/(M k_B T)}$ which at $T=0.5$ K is 65 nm.

Since $R_0 \gg a_{osc} \gg \lambda_B$, we can apply the local density or semiclassical approximation (LDA) \cite{pethick}. Here all thermodynamic quantities are function of the spatial coordinate. Furthermore, for temperatures not too close to $T_c$, the Thomas-Fermi approximation \cite{pethick} where the kinetic energy of the particles in the condensate is neglected, represents a rather good description of the condensate because the thickness of the layer where it breaks down  $\delta=\left(  a_{osc}^4 / R_0 \right)^{\nicefrac{1}{3}}$ \cite{pethick} is only $\delta = 0.15$ \textmu m. Therefore, the luminescence spectrum will be derived under these two approximations in the following.


It is by now well established that, at densities far below the Mott density, 
excitons can be described as a weakly interacting Bose gas \cite{okumura2001,zimmermann,semkat2009}. The interaction can be parametrized by a scattering length $a_S$ of the order of the Bohr radius, its magnitude depending on the details of the spin structure of the exciton states. 

We confine the excitons with an external potential $V_{\rm ext}(\vc{r})$, which for simplicity will be assumed to be that of a 3d harmonic oscillator $V_{\rm ext}=\alpha r^2$. 
The statistical theory of such a weakly interacting system 
with interaction energy $U_0= 4 \pi a_S \hbar^2/M$ in a potential trap  is well established for the atomic case \cite{griffin1996,bergeman2000,pethick} and we will shortly review these results, only.
In the standard mean-field theory, one has to solve the Gross-Pitaevski (GP) equation
which reads for $T=0$
\begin{equation}\label{eqn:gross}
-\frac{\hbar^2}{2M} \Delta \Psi (\vc{r})+ V(\vc{r})\Psi(\vc{r})+U_0 |\Psi(\vc{r})|^2 \Psi(\vc{r})= \mu \Psi(\vc{r})
\end{equation}
to obtain the wave function of the condensate $\Psi(\vc{r})$ and the condensate density $n_c(\vc{r})=|\Psi(\vc{r})|^2$ for a given chemical potential $\mu$.
 
The linear response and thus the luminescence spectrum is represented by the Hartree-Fock-Bogoliubov equations which we use in the Popov approximation (HFBP), which is valid also at temperatures around $T_c$.

As discussed earlier, we can apply the local density or semiclassical approximation (LDA). Here the densities of the condensate $n_c(\vc{r})$ and of the thermal excitons in excited states $n_T(\vc{r})$ and the Bogoliubov amplitudes are local functions $u(\vc{p},\vc{r})$, $v(\vc{p},\vc{r})$ that solve the coupled equations
\begin{equation}
\left( \begin{array}{c c}
{\cal L}(\vc{p},\vc{r}) & U_0 n_c(\vc{r})\\
-U_0 n_c(\vc{r}) & -{\cal L}(\vc{p},\vc{r})  
\end{array} \right)
\left( \begin{array}{c}
u(\vc{p},\vc{r})\\
v(\vc{p},\vc{r})  
\end{array} \right)
= \epsilon (\vc{p},\vc{r})
\left( \begin{array}{c}
u(\vc{p},\vc{r})\\
v(\vc{p},\vc{r})  
\end{array}
\right)
\end{equation}
 with 
 \begin{equation}
 {\cal L}(\vc{p},\vc{r})=p^2/2 M +V_{\rm ext}(\vc{r}) - \mu +2 U_0 n(\vc{r})
 \end{equation}
 and the renormalized energies  of the excited states  
 \begin{equation}
 \label{eqn:energy}
 \epsilon(\vc{p},\vc{r})=\left[({\cal L}(\vc{p},\vc{r}))^2- (U_0 n_c(\vc{r}))^2 \right]^{\nicefrac{1}{2}}\,.
 \end{equation}
 $n= n_c+n_T$ is the total density. The Bogoliubov amplitudes $u$ and $v$ are given by the usual relations 
\begin{subequations}
\begin{align}
\label{eqn:bogoliubov}
u(\vc{p},\vc{r})^2 &= \frac{1}{2} (({\cal L}(\vc{p},\vc{r})/\epsilon (\vc{p},\vc{r})+1)\,, \\
v(\vc{p},\vc{r})^2 &= \frac{1}{2} (({\cal L}(\vc{p},\vc{r})/\epsilon (\vc{p},\vc{r})-1)\,.
\end{align}
\end{subequations}
The density of the excitons in thermally excited states can be found by integrating over the excited states
\begin{eqnarray}
\label{eqn:thermal_density}
n_T(\vc{r})&=& \int \frac{d^3\vc{p}}{8 \pi^3}\left[ \frac{{\cal L}(\vc{p},\vc{r})}{\epsilon(\vc{p},\vc{r})}\left( 
n_B(\vc{p},\vc{r})+\frac{1}{2}\right) -\frac{1}{2}\right]\\
&& \times \Theta(\epsilon(\vc{p},\vc{r})^2) \nonumber
\end{eqnarray}
where the Bose function is given by
\begin{equation}
\label{eqn:bose}
n_B(\vc{p},\vc{r})= \frac{1}{\exp[\epsilon(\vc{p},\vc{r})/k_B T]-1} 
\end{equation}
and $\Theta$ is the Heaviside function which is equal to one when the argument is positive and zero otherwise.
For temperatures not too close to $T_c$, the thickness of the surface of the BEC cloud is much smaller than the radius $\delta \ll R_0$. Then one can neglect the kinetic energy term in Eq.\ (\ref{eqn:gross}) and the system can by be described quite accurately in the Thomas-Fermi approximation. Then the density of the condensate is given by \cite{bergeman2000}
\begin{equation}
\label{eqn:condensed_density}
n_c(\vc{r})= \frac{\mu - V_{\rm ext}(\vc{r}) -2 U_0 n_T}{U_0}\Theta(\mu - V_{\rm ext}(\vc{r}) -2 U_0 n_T)\,.
\end{equation}

For a given chemical potential, Eqs.\ (\ref{eqn:thermal_density}) and (\ref{eqn:condensed_density}) allow a self-consistent solution for all the relevant quantities.
The total number of particles $N$ is then found by integrating the total density over the volume of the trap. 

\begin{figure} [h]
\includegraphics[scale=0.35]{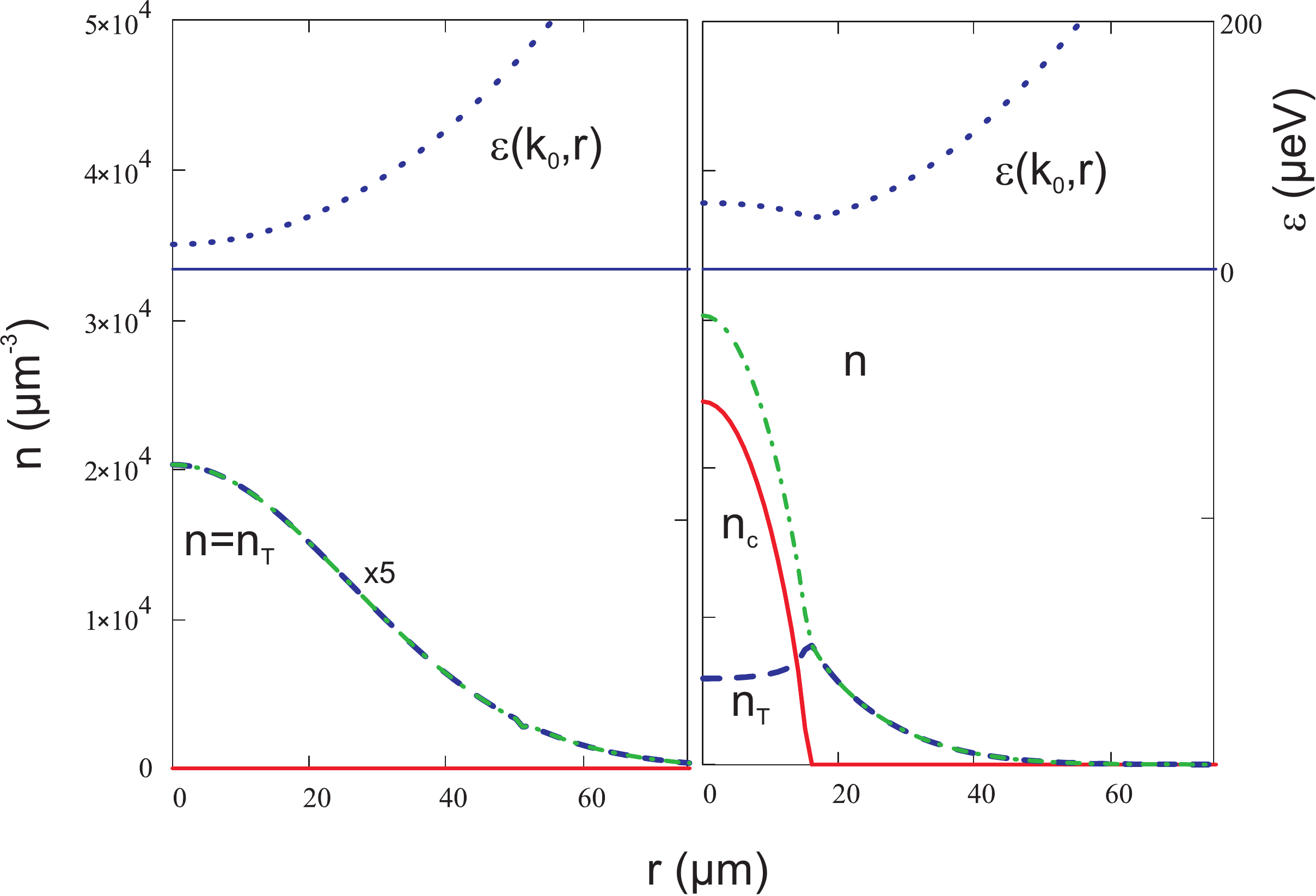}
\caption{Density (full and dashed lines) and renormalized quasiparticle energy (dotted line) for $N=1 \cdot 10^9$ excitons in the trap. $T=1$ K (left) and $T=0.5$ K (right). Parameters are: $\alpha=0.06$  \textmu eV/\textmu m$^2$, $U_0= 0.75$ neV/\textmu m$^3$. }
\label{fig:1}
\end{figure}

Finally, it should be noted that if the temperature is too high to allow for a condensate, the HFBP approximation goes over smoothly into the description of a weakly interacting Bose gas with a chemical potential $\mu = \mu_{\rm ideal} + 2 U_0 n_T$ \cite{bergeman2000}.
In Fig. \ref{fig:1} typical results for the density profiles obtained by this procedure are shown for the case of a normal system above $T_c$ (left) and with a condensate present (right). The calculation shows that the diameter of the cloud is somewhat smaller than predicted by the simple approximation given above (18 \textmu m vs.\ 24.5 \textmu m). 

Excitonic systems have one distinct property compared to other Bose gases in that they decay by emitting photons under energy and momentum conservation. This can proceed either directly, whereby momentum conservation requires that only excitons with the same momentum as the emitted photons are involved, or with assistance of momentum supplying phonons such that all exciton states can participate in the optical emission \cite{excitons}. The latter process has been considered already for a homogeneous Bose gas of interacting excitons by several authors \cite{shi1994,haug1983}.
Here it was shown, that the luminescence spectrum is determined by the excitonic spectral function $A(\vc{k},\omega)$
\begin{eqnarray}
\label{eqn:spectrum}
I(\omega^\prime) & \propto & 2 \pi |S(\vc{k}=0)|^2 \delta(\hbar\omega^\prime - \mu) n_c \\
&& + \sum_{\vc{k}\ne 0}|S(\vc{k})|^2n_B(\hbar\omega^\prime - \mu)A(\vc{k}, \hbar\omega^\prime - \mu) \nonumber
\end{eqnarray}
with $S(\vc{k})$ representing the exciton-photon coupling and $n_B$ being the usual Bose function (\ref{eqn:bose}). In the case of phonon-assisted transitions, we have $\omega^\prime = \omega - \omega_{gX} -\omega_{\rm phonon}$ with $\hbar \omega_{gX}$ being the excitonic band gap of the semiconductor. $S(\vc{k})$ can be assumed to be $\vc{k}$ independent. The first term in Eq.\ (\ref{eqn:spectrum}) gives rise to a $\delta$ shaped luminescence line at the position of the chemical potential of the system, the strength of which is determined by the coupling function at $k=0$ and the condensate density.

The case of the 3d bulk state with direct exciton-photon coupling can be treated in the same way with $\omega^\prime = \omega - \omega_{gX}$ and $S(\vc{k})= S_0 \delta(\vc{k}-\vc{k}_0)$. 
$\vc{k}_0$ is the wavevector of the intersection of photon and exciton dispersion. Its modulus is given by $k_0= \omega_{gX} n/c$, where $n$ is the refraction index and $c$ is the vacuum velocity of light.
Here we see already a remarkable difference between direct and phonon-assisted luminescence processes: due to the form of $S(\vc{k})$, the condensate itself will not be detectable by the direct luminescence process!

A first principle calculation of the decay luminescence spectrum in a trap is a challenging task (see, e.g., \cite{zimmermann} for the case of a 2d potential trap). Here we proceed in a much simpler way by noting that the optical wavelength of the emission (in case of Cu$_2$O about 200 nm) is much smaller than the size of the exciton cloud with diameter $2 R_0$. Therefore, one can apply a local approximation also for the spectral function, which then becomes that of the homogeneous case \cite{shi1994} but now in addition a function of $\vc{r}$:
\begin{eqnarray}
\label{eqn:spectral_function}
\lefteqn{A(\vc{r},\vc{k},\omega)=} \\
&& 2 \pi\left[ 
u(\vc{k,\vc{r}})^2 \delta(\hbar \omega-\epsilon(\vc{k},\vc{r})) -v(\vc{k},\vc{r})^2 \delta(\hbar \omega+\epsilon(\vc{k},\vc{r}))
\right]\,.\nonumber
\end{eqnarray}
Here $\epsilon(\vc{k},\vc{r})$ are the renormalized energies (\ref{eqn:energy}).
This means that in Eq.\ (\ref{eqn:spectrum}) the frequency $\omega^\prime$ is determined by the local exciton energy  $\hbar\omega^\prime_{loc} = \hbar\omega- \hbar\omega_{gX}-V_{\rm ext}(\vc{r})$ while $\mu$ is the local chemical potential $\mu_{loc}= \mu-V_{\rm ext}(\vc{r})$. Obviously, the external potential cancels in the argument of $I(\omega^\prime)$, which depends only on the global chemical potential of the system, as it should. The lineshape of the luminescence spectrum is determined by the renormalized energies of the excited states of the system, but now evaluated at each point in the trap. 

While for the phonon-assisted process, Eq.\ (\ref{eqn:spectral_function}) gives rise to a smooth spectrum \cite{shi1994}, the direct process behaves differently. Emission will come only from the states with wavevector $k_0$. The intensity, therefore, reflects only the occupation of this state, but the spectral position of the line at $\mu + \epsilon(\vc{k}_0,\vc{r})$  directly gives the renormalization of the quasiparticle energy dispersion due to the condensate. Furthermore, due to the pole of the $\delta$-function at $\hbar \omega^\prime =\mu  -\epsilon(\vc{k},\vc{r})$ in the condensate ($v\ne 0$) emission occurs also at the low energy side of the chemical potential. Both effects provide unique and sensitive footprints of the onset of BEC.


We now apply the foregoing results to study the behavior of the luminescence line in the case of a weak direct exciton-photon interaction with parameters adjusted to the case of the lowest exciton state in Cu$_2$O.
Made up from both positive parity and doubly degenerate valence and conduction bands, the four exciton states 
split in a triply degenerate orthoexciton and in a single paraexciton, which is the energetically lowest exciton state, split off by 12 meV from the ortho states due to electron-hole exchange \cite{snoke2000}. While the latter are optically 
weakly allowed (quadrupole transition, oscillator strength $3\cdot 10^{-9}$ \cite{froehlich1991}),
the paraexciton as a pure spin triplet state is forbidden in all orders. Its intrinsic decay is only possible via an 
odd parity optical phonon with $\Gamma^-_5$ symmetry, from which we expect a very long lifetime of these exciton states \cite{brandt2007} and thus an almost true equilibrium BEC of 3d excitons.

\begin{figure*} [t]
\includegraphics[scale=0.4]{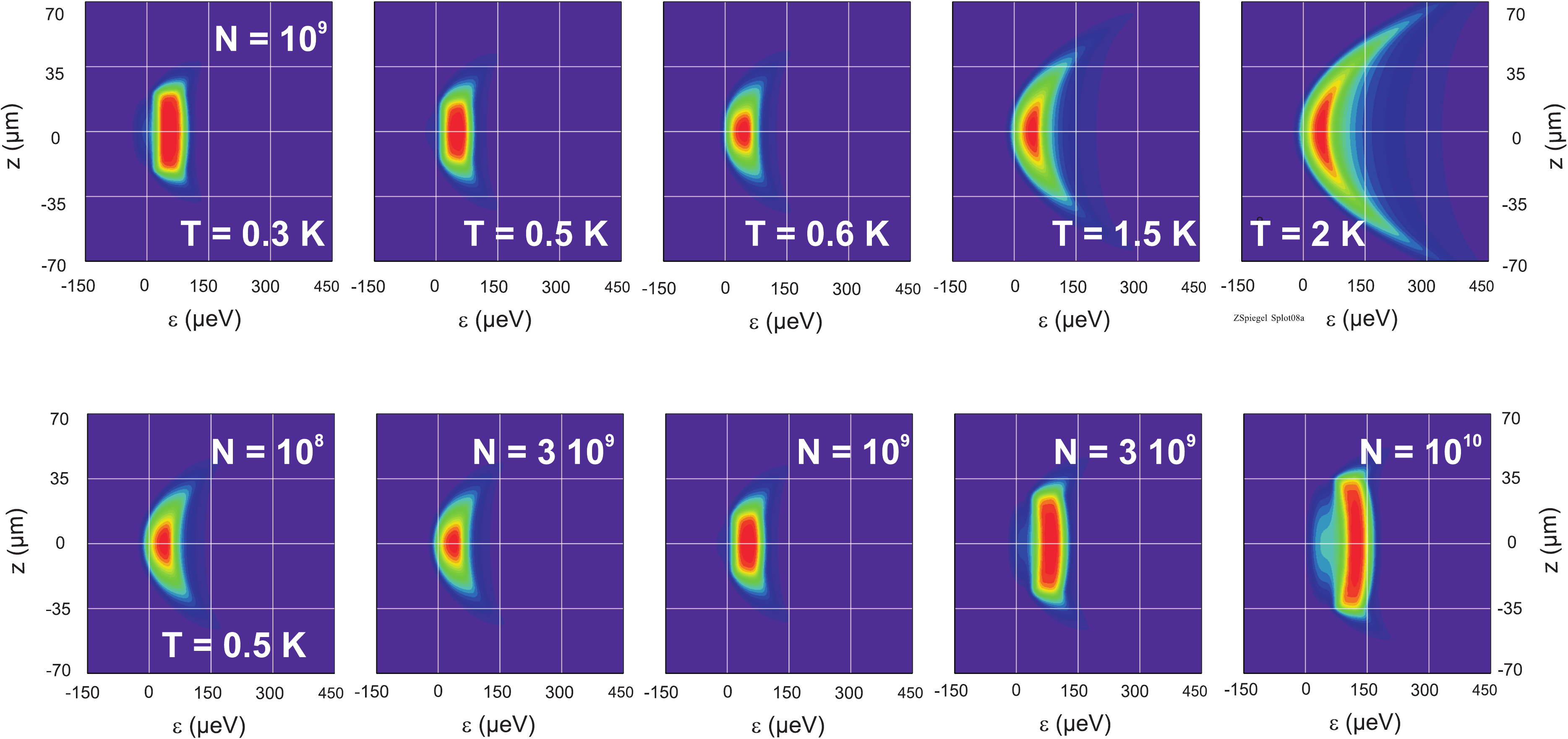}
\caption{Luminescence spectra of the direct luminescence process. Upper row with constant particle number $N=1 \cdot 10^9$ and at constant temperature ($T=0.5$ K) with increasing particle numbers in the trap (lower row).}
\label{fig:2}
\end{figure*}

In  a strain trap, the paraexciton becomes weakly allowed due to mixing with higher lying exciton states, the oscillator strength remains quite small so that the theory given above is applicable. The weak variation of the transition probability with strain across the trap will be neglected. 
In the typical experimental situation, one images a small stripe of width $\Delta x$  elongated along the $z$ direction centered in the center of the trap onto the entrance slit of a spectrograph. Integrating over the $y$-direction perpendicular to $z$ we obtain a spatially resolved spectrum $I(z,\omega)$. In order to compare with any real experimental situation, one has to convolute the spectra with the finite resolution of the spectrograph. For this we take a slit function of supergaussian shape $s(x)\propto \exp(-(x/\Delta)^4)$ with FWHM $1.825 \Delta = 75$ \textmu eV. 

There are three parameters that influence the behavior of the excitons: trap potential constant $\alpha$, interaction strength $U_0$ and mass $M$. For the model calculations we assumed the following parameters $M=2.6 m_e, \alpha = 0.06$ \textmu eV/\textmu m$^2$ and $U_0= 0.75$ neV \textmu m$^3$. In Fig.\ \ref{fig:2} we have plotted a series of spatially resolved spectra for a range of exciton numbers and temperatures. While the upper row shows the variation with temperature at constant $N=10^9$, the lower series demonstrates the influence of $N$ at a temperature of $T=0.5$ K. 
While at $T>T_c$ or equivalently $N<N_c$, the lineshape follows strictly the parabolic shape of the potential well, as one expects for a normal gas, below $T_c$ the spectrum changes drastically. The low energy side becomes almost flat. For very high particle number a weak shoulder develops at the low energy side, which represents the anomalous luminescence via the negative pole due to the condensate. In contrast to the case of a noninteracting system, the spatial width of the spectrum may become even larger than the thermal width (compare, e.g., spectra for $N=3\cdot 10^9$ and $N=10^{10}$).
The quantitative behavior, of course, will depend on the fine details of the shape of the potential trap, but the qualitative features will be the same. Thus, these drastic changes in the luminescence spectrum can be considered as a unique footprint of the Bose-Einstein condensation of excitons in a potential trap. 

Finally, we ask whether an experimental realization seems to be possible with the present knowledge of the exciton properties in Cu$_2$O. Previous experiments with excitons in Cu$_2$O \cite{wolfe1986,wolfe1993,snoke2000b} have shown that, under quasi-cw excitation with an absorbed laser power of 50 mW, $2\cdot 10^{9}$ excitons can be put into a trap but with an excitonic temperature of about 2.5 K which certainly is not enough for a BEC. Extrapolating the data, one should either increase the pump power to a value of 10 W or reduce the temperature of the exciton system below 0.75 K. Both strategies seem to be possible by present day technology. 
   
We have shown that the luminescence spectrum of the direct recombination luminescence of excitons in a potential trap changes in a unique way if a condensate of excitons is present. This change is independent of the details of the excitonic systems and reflects directly the renormalization of the quasiparticle energies due to the interaction of the excitons. 

We acknowledge the support by the Deutsche For\-schungsgemeinschaft (SFB 652 ``Starke Korrelationen im Strahlungsfeld'').


\end{document}